\documentclass[aps,prl,reprint,superscriptaddress]{revtex4-1}
\usepackage{color,graphicx}
\usepackage[normalem]{ulem}

\begin{document}


\title{X-Ray Photon Correlation Spectroscopy Reveals Intermittent Aging Dynamics in a Metallic Glass}


\author{Zach Evenson}
 \email{Zachary.Evenson@frm2.tum.de}
\affiliation{Heinz Maier-Leibnitz Zentrum (MLZ) and Physik Department, Technische Universit\"{a}t M\"{u}nchen, Lichtenbergstrasse 1, 85748 Garching, Germany}
\affiliation{Institut f\"{u}r Materialphysik im Weltraum, Deutsches Zentrum f\"{u}r Luft- und Raumfahrt (DLR), 51170 K\"{o}ln, Germany}

\author{Beatrice Ruta}
\affiliation{ESRF - The European Synchrotron, CS40220, 38043 Grenoble, France}

\author{Simon Hechler}
\affiliation{Chair of Metallic Materials, Department of Materials Science and Engineering, Saarland University, Campus C6.3, 66123 Saarbr\"{u}cken, Germany}

\author{Moritz Stolpe}
\affiliation{Chair of Metallic Materials, Department of Materials Science and Engineering, Saarland University, Campus C6.3, 66123 Saarbr\"{u}cken, Germany}

\author{Eloi Pineda}
\affiliation{Department of Physics, Universitat Polit\`{e}cnica de Catalunya-BarcelonaTech, Esteve Terradas 8, Castelldefels 08860, Spain}

\author{Isabella Gallino}
\affiliation{Chair of Metallic Materials, Department of Materials Science and Engineering, Saarland University, Campus C6.3, 66123 Saarbr\"{u}cken, Germany}

\author{Ralf Busch}
\affiliation{Chair of Metallic Materials, Department of Materials Science and Engineering, Saarland University, Campus C6.3, 66123 Saarbr\"{u}cken, Germany}


\date{\today}

\begin{abstract}

We use coherent X-rays to probe the aging dynamics of a metallic glass directly on the atomic level. Contrary to the common assumption of a steady slowing down of the dynamics usually observed in macroscopic studies, we show that the structural relaxation processes underlying aging in this metallic glass are intermittent and highly heterogeneous at the atomic scale. Moreover, physical aging is triggered by cooperative atomic rearrangements, driven by the relaxation of internal stresses. The rich diversity of this behavior reflects a complex energy landscape, giving rise to a unique type of glassy-state dynamics.
\end{abstract}


\maketitle

\setlength{\parskip}{0mm}

	Physical aging is not only of scientific interest \cite{Struik1978,Berthier2011,*Cipelletti2003} but also of great practical importance, as the performance stability of many technologically relevant amorphous materials such as oxide glasses, polymers and metallic glasses (MGs) depends on how aging affects their properties during the expected service life. For MGs in particular, aging can result in severe embrittlement \cite{Murali2005} and bring about profound changes to fracture and fatigue properties \cite{Launey2007,*Launey2008}. Therefore, the ability to understand and characterize the micromechanisms of aging in MGs will play a central role in ensuring their success in a broad range of industrial and commercial applications. 

	A key to understanding aging on the microscopic level is measuring the time scale on which the system rearranges its internal structure. These structural rearrangements are directly related to how fast microscopic density fluctuations in the system decay \cite{Boon1980}. The physical quantity describing this is the density correlation function, which can be measured over a wide range of time scales using a variety of complimentary experimental techniques \cite{Mezei1987,*DeGennes1959,*Madsen2014}. Additionally, it can be formally defined within the framework of statistical mechanics, allowing for detailed comparisons between experiment, theory \cite{Gotze1999} and simulation \cite{Kob1995}. Access to the density correlation function is therefore fundamental for describing the dynamical behavior of glasses and achieving a consistent picture of structural relaxation on microscopic length scales.
	
	X-ray photon correlation spectroscopy (XPCS) has recently emerged as a novel technique for studying the microscopic dynamics of condensed matter \cite{Ruta2012,Madsen2014,Leitner2009,Leitner2012}. In XPCS, coherent beams of X-rays are scattered from a sample and the resulting fluctuations in intensity correlated over time. This enables measurement of the density correlation function on the atomic scale. Previous XPCS investigations of hyper-quenched Mg$_{65}$Cu$_{25}$Y$_{10}$ and Zr$_{66}$Ni$_{33}$ MGs \cite{Ruta2012,Ruta2013} have revealed the presence of a unique stress-dominated dynamics characterized by subsequent aging regimes: a fast exponential growth for short waiting times, followed by a long, almost stationary regime. The latter has been observed also in network glasses \cite{Ruta2014a} and is contrary to the steady slowing down of the dynamics during aging observed in macroscopic studies.
	
\setlength\belowcaptionskip{-3ex}

\begin{figure}
\includegraphics[width=0.50\textwidth]{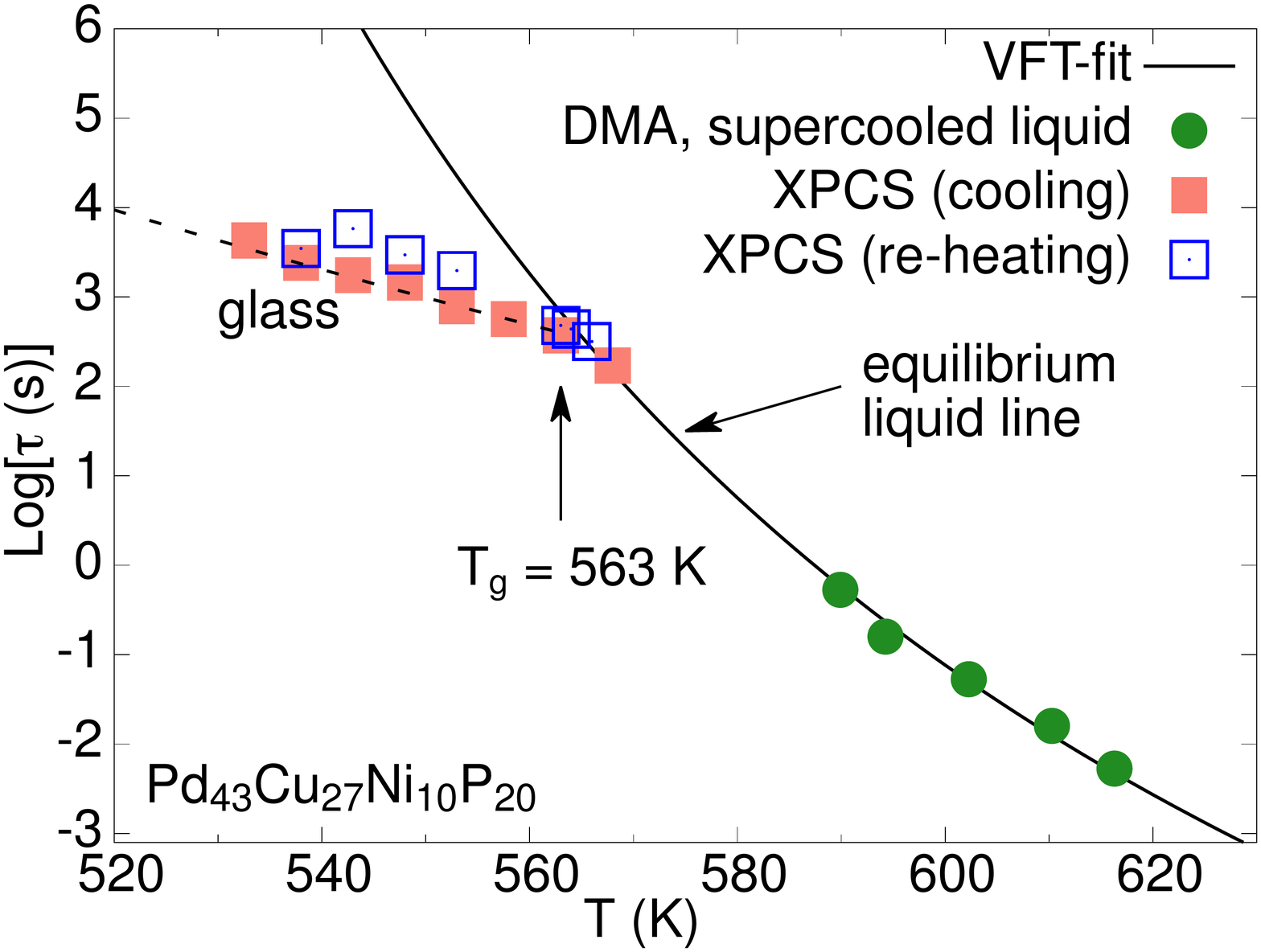}%
\caption{\label{fig:relax} Structural relaxation times measured with DMA (filled circles) and XPCS during slow cooling from the supercooled liquid (filled squares) and subsequent re-heating from the partially annealed glass (open squares). The solid line is a fit of the equilibrium data to the VFT equation. The dashed line is an Arrhenius fit to the XPCS cooling data in the glass.}
\end{figure}			

	Here, we investigate the atomic motion in a Pd$_{43}$Cu$_{27}$Ni$_{10}$P$_{20}$ MG ribbon using XPCS and show that this system ages in a highly heterogeneous manner, consisting of quiescent periods of stationary dynamics interspersed with cooperative, avalanche-like structural rearrangements. Our findings suggest the existence of a complex dynamical scenario occurring at the atomic level and are consistent with a stress-dominated dynamics that controls aging. The XPCS measurements were performed at the beamline ID10 at ESRF, France. A detailed overview of the experimental set-up and data treatment procedure can be found in the Supplementary Material \cite{SuppMat_XPCS_PRL}.

	The structural relaxation times $\tau$ of Pd$_{43}$Cu$_{27}$Ni$_{10}$P$_{20}$ in the ultra-viscous supercooled liquid above $T_g$ were determined from dynamic mechanical analysis (DMA) measurements \cite{SuppMat_XPCS_PRL} employed at different test frequencies and are given by the filled circles in Fig. \ref{fig:relax}. We model the temperature dependence of $\tau$ by fitting the empirical Vogel-Fulcher-Tammann (VFT) equation $ \tau(T) = \tau_0\exp[D^*T_0/(T-T_0)]$ to the equilibrium data. Here $D^*$ is a measure of an inherent liquid-state property, the so-called fragility \cite{Angell1995}. This reflects the degree of departure of the liquid dynamics from the Arrhenius equation, with smaller values of $D^*$ corresponding to more fragile liquids. We find $D^*$ = 10.3 $\pm$ 0.4, in agreement with previously reported rheological data \cite{Fan2004a}.

	The square symbols in Fig. \ref{fig:relax} are the values of $\tau$ measured using XPCS. Below $T_g = 563$ K, the transition into the out-of-equilibrium glass is apparent. In the glass $\tau$ displays a much weaker temperature dependence than in the supercooled liquid which is well described by an Arrhenius equation with an activation energy of 178.9 $\pm$ 0.4 kJ/g-atom. The open squares represent $\tau$ measured during subsequent reheating of the partially aged glass from 533 K. Relaxation in the glass depends on the previous thermal history, reflected here by the higher values of $\tau$ as a result of the sample aging.
	
\begin{figure}
\includegraphics[width=0.4\textwidth]{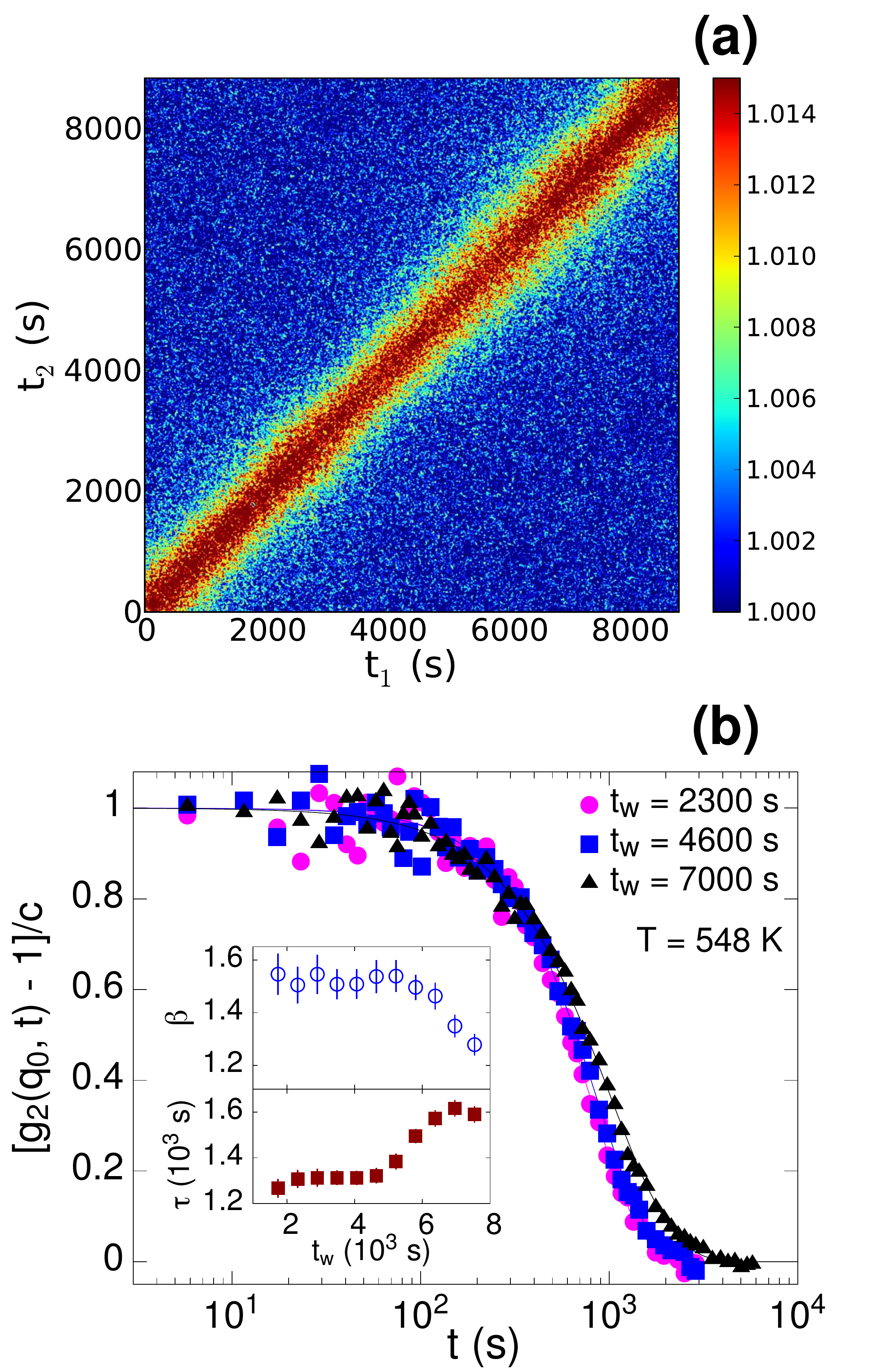}%
\caption{\label{fig:stationary}(a) Two-time correlation function (TTCF) at T = 548 K, measured in the glass upon step-cooling from the supercooled liquid. The width of the diagonal intensity profile is proportional to $\tau$. (b) Normalized $g_2(q_0 = 2.8$ \AA$^{-1}, t)$ extracted from the TTCF after different waiting times $t_w$. Solid lines through the data are best-fits to the KWW equation. The values of $\tau$ and $\beta$ are given in the inset as a function of $t_w$. A stationary dynamics persists up to $4.6 \times 10^3$ s, after which the dynamics slows down as sample ages.}
\end{figure}

	The time evolution of the microscopic dynamics during aging is directly captured in XPCS by the intensity two-time correlation function $G(q,t_1,t_2)$ (TTCF). This quantity reflects the statistical similarity between average configurations measured at times $t_1$ and $t_2$ and is thus the time-resolved version of the standard intensity correlation function $g_2(q, t) = \langle G\rangle$ \cite{Madsen2014}. 	The TTCF measured upon cooling from the supercooled liquid during an isothermal hold of 2.2 hr at 548 K ($T_g - 15$ K) is shown in Fig. \ref{fig:stationary}(a). Along the center diagonal, which corresponds to the elapsed time of the measurement, the values of the TTCF are largest and decrease as the difference between $t_1$ and $t_2$ grows. The width of the diagonal contour is thus directly proportional to $\tau$, i.e. the characteristic time scale on which any given atomic configuration no longer corresponds to that measured at a later time.
	
	The normalized $g_2(q_0, t)$ function, extracted from the TTCF after progressively longer waiting times $t_w$, is shown in Fig. \ref{fig:stationary}(b). Each correlation function was fitted using a Kohlrausch-Williams-Watts (KWW) expression $g_2(q_0, t) = 1 + c\exp[-2(t/\tau)^{\beta}]$, where $\tau$ is the relaxation time, $\beta$ is the shape parameter \cite{Martin2000} and $c = \gamma f_q^2$ is the product between the experimental contrast $\gamma$ and the square of the Debye-Waller factor $f_q$. The temporal dependence of $\tau$ and $\beta$ is displayed in the inset.
	
	Generally, aging is expected to begin as soon as the sample falls out of equilibrium and $\tau$ becomes comparable to the experimental observation time \cite{Hodge1994}. Surprisingly, we observe the onset of aging here only after a long stationary regime of $4.6 \times 10^3$ s [Fig. \ref{fig:stationary}(b)]. A similar stationary dynamics has been reported in previous XPCS studies of as-quenched MGs \cite{Ruta2012,Ruta2013}. In those works, however, stationary dynamics was only observed following an initial, fast aging regime, while the measurements presented in Fig. \ref{fig:stationary} show the opposite trend. Nonetheless, in the potential energy landscape (PEL) approach \cite{Doliwa2003,*Heuer2008}, both stationary dynamics suggest the presence of deep energy basins, in which the system becomes trapped and explores local minima separated by similar energy barriers. During aging, the system eventually escapes the basin and begins to explore deeper energy minima. The discrepancy in the occurrence of our stationary dynamics and that reported in Refs. \cite{Ruta2012,Ruta2013} is likely the simple consequence that our glass is studied upon slow cooling from the supercooled liquid, whereas those studies had measured hyper-quenched ribbons trapped already at much higher levels on the PEL.

	These observations are consistent with recent molecular dynamics (MD) simulations of a Cu$_{56}$Zr$_{44}$ MG by Fan \textit{et al.}, which show that the microscopic relaxation dynamics is significantly more spatially localized in a relaxed or slowly cooled glass, than in a fast quenched material \cite{Fan2015}. 
	
\begin{figure}
\includegraphics[width=0.4\textwidth]{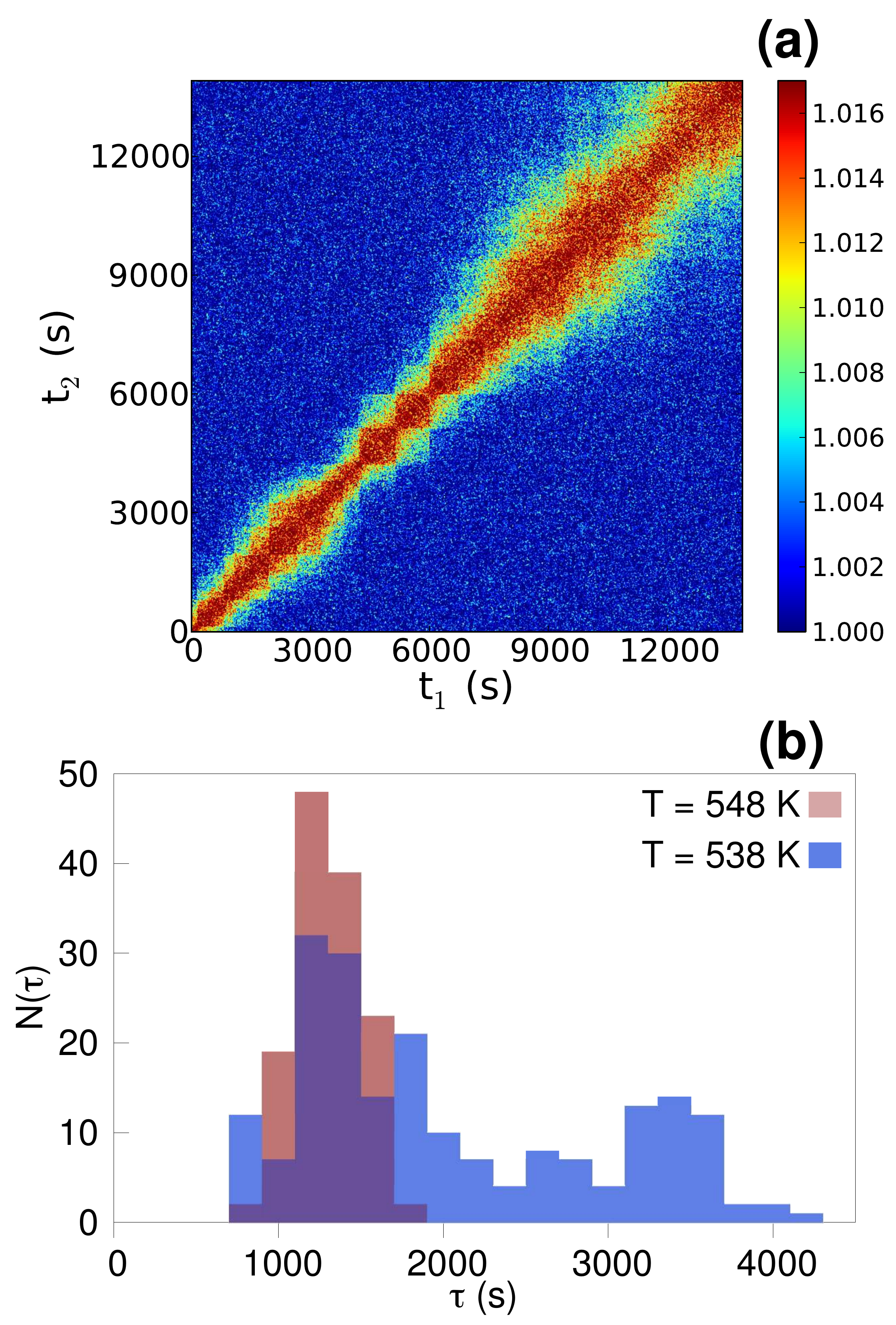}%
\caption{\label{fig:hetero}(a) Measured TTCF at T = 538 K after step-cooling from 548 K. The overall broadening of the intensity profile reflects a slowing down of the microscopic dynamics corresponding to physical aging. (b) Distribution of $\tau$ during stationary dynamics (T = 548 K) and heterogeneous aging dynamics (T = 538 K).}
\end{figure}		
	
	The peculiar dynamical scenario observed at 548 K becomes even more dramatic at lower temperatures, where a highly heterogeneous relaxation dominates the observed aging behavior. Figure \ref{fig:hetero}(a) shows the TTCF measured after step-cooling from 548 to 538 K and holding isothermally for 3.9 hr. The broadening trend of the intensity profile with increasing $t_w$ clearly shows that the dynamics is slowing down significantly during annealing. A remarkable feature, however, is that the aging process cannot be solely described by a more or less steady slowing down of the dynamics, as observed in previous XPCS investigations \cite{Ruta2012,Ruta2013}. Instead, rapid decorrelation events -- characterized by a sudden narrowing of the TTCF intensity profile -- appear intermittently as the measurement progresses. A closer inspection of $g_2(q_0,t)$ as a function of $t_w$ also reveals the presence of multiple stationary regimes \cite{SuppMat_XPCS_PRL}. The histograms in Fig. \ref{fig:hetero}(b) show the distribution of $\tau$ at both 548 and 538 K, obtained from fitting the KWW-equation to the extracted $g_2(q_0,t)$ \cite{SuppMat_XPCS_PRL}. The stationary dynamics at 548 K (Fig. \ref{fig:stationary}) is shown by the unimodal behavior of $\tau$ around $1.5 \times 10^3$ s, while the heterogeneous aging at 538 K  exhibits a bimodal distribution. The appearance of a larger value of $\tau \sim 3 \times 10^3$ s at 538 K reflects the presence of the noticeably slower relaxation connected with the aging process.

	Since XPCS measures the contribution from all components of the system, according to their characteristic time scale of motion, it might be speculated if this behavior is attributed to the decoupling of individual component diffusivities, as observed in the supercooled liquid by Bartsch \textit{et al.} \cite{Bartsch2010}. We rule out this hypothesis, as those self-diffusion data could lead only to a collective dynamics given by the average of distinct correlation functions with decay times very far from one another, and not to a temporal intermittent dynamics.  The avalanche-like and intermittent temporal fluctuations in the collective dynamics of Fig. \ref{fig:hetero}a emanate instead from distinct changes in the decay of correlations at the atomic length scale and thus contradict our intuition of a steady slowing down of the glass's internal clock during aging. Intermittent dynamics, however, was not observed in earlier XPCS measurements of a Mg$_{65}$Cu$_{25}$Y$_{10}$ MG following a similar thermal protocol \cite{Ruta2012}. This suggests a possible connection to the higher fragility of the Pd$_{43}$Cu$_{27}$Ni$_{10}$P$_{20}$ supercooled liquid \cite{Mauro2008,*Wei2014}, although further experiments on other MG systems would be necessary to establish a definitive relationship.
	
\begin{figure*}
\includegraphics[width=0.7\textwidth]{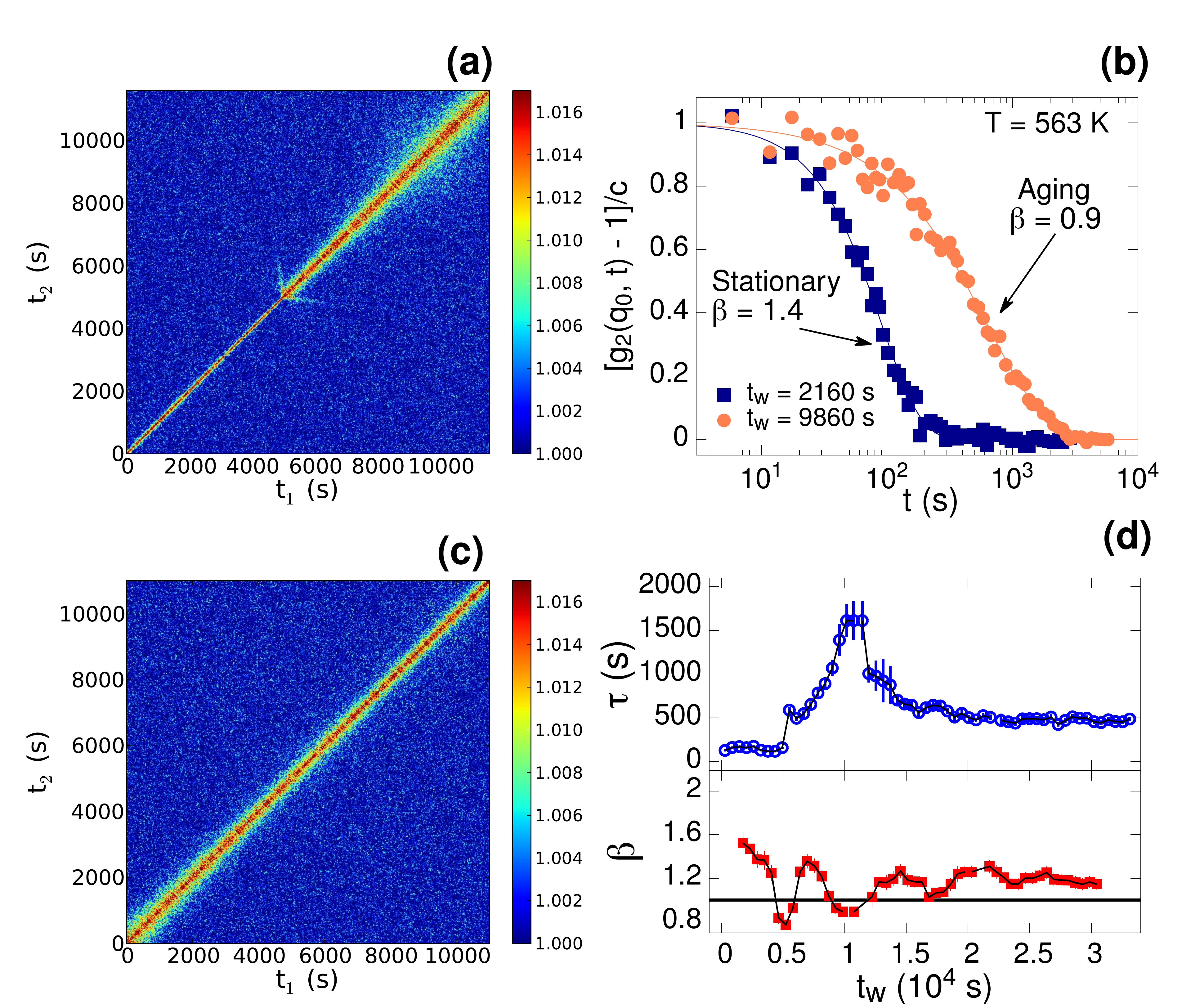}%
\caption{\label{fig:fourpanel}(a) Measured TTCF after directly heating the annealed sample to T = 563 K. (b) Normalized one-time correlation functions calculated in the initially stationary (filled squares) and aging (filled circles) regime. (c) TTCF corresponding to the continuation of the measurement at T = 563 K. (d) Evolution of $\tau$ and $\beta$ over the entire length of the measurement.}
\end{figure*}	

	In Fig. \ref{fig:fourpanel} the evolution of the TTCF is monitored after heating the now partially aged sample directly to 563 K \cite{SuppMat_XPCS_PRL}. A long regime of fast, stationary dynamics is observed up $t_w = 5 \times 10^3$ s, which is interrupted by a sudden, then gradual broadening of the TTCF up to $t_w = 1.1 \times 10^4$ s [Fig. \ref{fig:fourpanel}(a)]. The $g_2(q_0,t)$ calculated in the initial stationary regime displays a fast dynamics with $\tau = 150$ s and a compressed exponential shape parameter of $\beta = 1.4$ [Fig. \ref{fig:fourpanel}(b)]. Afterwards, the dynamics has slowed considerably as the sample ages ($\tau = 960$ s), accompanied by a drop in the value of $\beta$ to 0.9. During longer annealing we observe the dynamics to speed up ($\tau = 500$ s) and the shape of the correlation function becomes re-compressed with $\beta = 1.2$ [Fig. \ref{fig:fourpanel}(c)-\ref{fig:fourpanel}(d)]. 

	The unique compressed shape of the correlation functions has been linked to ballistic-type motions that arise from the presence of internal stresses stored into glassy (or jammed) states during structural arrest \cite{Ruta2012,Ruta2013,Cipelletti2003,Bouchaud2001,*Ferrero2014}. This is in contrast to diffusive processes, characteristic of mass transport in liquids, which are driven by thermal energy and result in simple or stretched ($\beta < 1$) exponential decays \cite{Meyer2010,*Yang2011}. By choosing $\beta$ as an \textsl{ad-hoc} metric to reflect the evolution of microscopic stresses in the sample during aging, we can thus relate the sudden increase in $\tau$ and concomitant decrease in $\beta$ observed in these measurements (see also \cite{SuppMat_XPCS_PRL}) to transitions from more-stressed to less-stressed states as the sample ages \cite{Cipelletti2003}. In fact, the decrease in $\beta$ from 1.5 to $\sim$ 1.2 [inset, Fig. \ref{fig:stationary}(b)] agrees very well with the short- and the long-time values, respectively, predicted by the model of Bouchaud and Pitard explaining the compressed decay of the correlation functions in terms of a release of internal stresses \cite{Bouchaud2001}. In this sense, we envision aging in MGs as the consequence of the gradual release of atomic-level stresses.

	A structural mechanism has been proposed that links these rearrangements to localized micro-collapses of groups of particles, which trigger subsequent collapses in neighboring regions through the formation of stress dipoles \cite{Bouchaud2001}. This avalanche-like dynamics has been found in a number of phenomena including martensitic transformations \cite{Muller2011,*Sanborn2011}, deformation of MGs \cite{Fan2015,Antonaglia2014,*Krisponeit2014}, crystallization of a hard-sphere glass \cite{Sanz2014} and shear flow of droplet emulsions through a thin opening \cite{Chen2011}. We therefore believe that these microscopic rearrangement events act as important mediators in the aging of MGs via the cooperative relief of atomic-level stresses. This point is illustrated in Fig. \ref{fig:fourpanel}, where our heating of the annealed sample to 563 K apparently reduces the degree of internal stresses, allowing the aging process to eventually re-activate.
	
	The transition in Fig. \ref{fig:fourpanel}b between stationary and aging dynamics is accompanied by a decrease in $\beta$ from 1.4 to 0.9, thus suggesting a transition from ballistic to diffusive, liquid-like particle motion. Additionally, the final value of $\beta = 1.2$ measured at 563 K indicates that the sample is still stuck in a non-equilibrium state. This is surprising, as the duration of the annealing performed at this temperature ($\sim 3 \times 10^4$ s) far exceeds the extrapolated macroscopic $\tau$ of the equilibrium liquid ($\sim 1 \times 10^3$ s). It would thus appear that -- at least in the microscopic sense -- full re-equilibration into the supercooled liquid is astonishingly difficult. This is further confirmation that macroscopic and microscopic aging in this MG clearly obey different laws.
	
	In closing, we note that similar aging dynamics has also been observed in colloidal gels and associated with intermittent ballistic particle motion \cite{Duri2006}. Our results thus strengthen the similarities between soft matter and MGs, as well as their differences with respect to network glasses, which are characterized by a stretched exponential decay of the density correlation function, even in the glass \cite{Ruta2014a}. The intermittent dynamics reported here for the first time in a structural glass is indicative of a complex energy landscape \cite{Doliwa2003,*Heuer2008}, characterized by numerous minima, in which the system dwells for long periods of time in stable configurations and reflect the presence of both localized and cascade relaxation dynamics, as in the simulations of Ref.  \cite{Fan2015}. In this Letter we have also shown that, during the long annealing at each temperature (which was not simulated in Ref. \cite{Fan2015}), we observe a spontaneous re-activation of the aging from a localized dynamical regime, not reported in previous works. Nevertheless, these results should provide the impetus for a deeper exploration of the connection between local features of the PEL and out-of-equilibrium dynamics by means of simulation efforts, as well as for the development of new ways to exploit additional information from XPCS experiments.

\begin{acknowledgments}
Karim L\'{}Hoste and Hugo Vitoux are gratefully acknowledged for their technical support during the XPCS measurements. It is a pleasure to thank Yuriy Chuskin for providing the code for the XPCS data treatment. Z.E. is supported by the German Academic Exchange Service (DAAD). E.P. is supported by MINECO, grant FIS2014-54734-P and Generalitat de Catalunya, grant 2014SGR00581.
\end{acknowledgments}

\bibliography{XPCS_Pd43}

\end{document}